# BIGP- A NEW SINGLE PROTOCOL THAT CAN WORK AS AN IGP (INTERIOR GATEWAY PROTOCOL) AS WELL AS EGP (EXTERIOR GATEWAY PROTOCOL)

Isha Gupta
*ASET/CSE Department, Noida, India*
*isha.gupta0701@gmail.com*

*Abstract-* EGP and IGP are the key components of the present internet infrastructure. Routers in a domain forward IP packet within and between domains. Each domain uses an intra-domain routing protocol known as Interior Gateway Protocol (IGP) like IS-IS, OSPF, RIP etc to populate the routing tables of its routers. Routing information must also be exchanged between domains to ensure that a host in one domain can reach another host in remote domain. This role is performed by inter-domain routing protocol called Exterior Gateway Protocol (EGP). Basically EGP used these days is Border Gateway Protocol (BGP). Basic difference between the both is that BGP has smaller convergence as compared to the IGP's. And IGP's on the other hand have lesser scalability as compared to the BGP. So in this paper a proposal to create a new protocol is given which can act as an IGP when we consider inter-domain transfer of traffic and acts as BGP when we consider intra-domain transfer of traffic.

## I. INTRODUCTION

Routing involves two basic activities: determination of optimal routing paths and the transport of packets through an internetwork. The transport of packets through an internetwork is relatively straightforward. Path determination, on the other hand, can be very complex. One protocol that addresses the task of path determination in today's networks is the *Border Gateway Protocol* (BGP). BGP performs inter-domain routing in Transmission-Control Protocol/Internet Protocol (TCP/IP) networks. BGP is an exterior gateway protocol (EGP), which means that it performs routing between multiple autonomous systems or domains and exchanges routing and reachability information with other BGP systems. BGP was developed to replace its predecessor, the now obsolete *Exterior Gateway Protocol* (EGP), as the standard exterior gateway-routing protocol used in the global Internet. BGP solves serious problems with EGP and scales to Internet growth more efficiently [1,11].

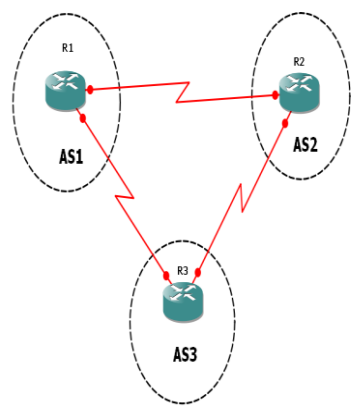

Figures 1: BGP topology

In the above figure a simple BGP topology is shown. It consists of three autonomous systems which are linked together using serial cables. R1, R2 and R3 are core routers of the autonomous system AS1, AS2 and AS3 respectively. The primary function of a BGP system is to exchange network-reachability information, including information about the list of autonomous system paths, with other BGP systems. Each BGP router maintains a routing table that lists all feasible paths to a particular network. The router does not refresh the routing table, however. Instead, routing information received from peer router is retained until an incremental update is received [2, 15 and 16].

IGP (interior gateway protocol) are the protocols which are used to route the data in an autonomous system [6, 8]. Three main branches of routing protocol algorithms exist for IGP routing protocol Distance vector, Link-state and Balanced hybrid. RIP was first popularly used IP distance vector protocol, with the cisco-proprietary Interior

Gateway Routing Protocol (IGRP) being introduced a little later. But because of their slow convergence and other problems Link-state protocols-in particular, OSPF and integrated IS-IS-solved the main issue with distance vector protocols [3, 6]. Around same time as the introduction of OSPF, cisco created a proprietary routing protocol called Enhanced Interior Gateway Routing Protocol (IGRP).

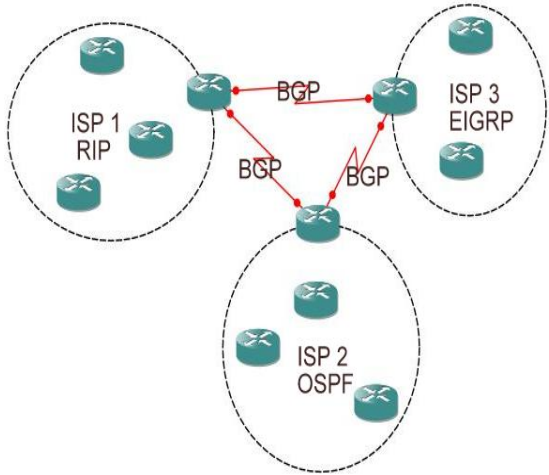

Figure 2: Comparing Locations used for implementing IGPs and EGPs.

In the above figure three different ISPs are shown. Different IGPs are used to communicate intra-ISP. ISP1, ISP2 and ISP3 using RIP, OSPF and EIGRP protocols respectively. Inter-ISP communication takes place using BGP. Here each ISP can be treated as a single autonomous system.

## II. HEADER OF BIGP

For inter and intra autonomous system interaction we have developed BIGP. Each packet will have a header of BIGP protocol as shown in figure below.

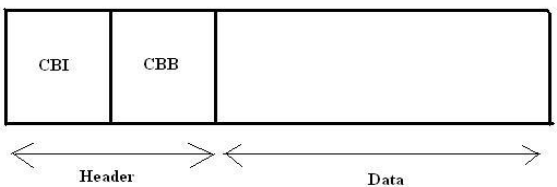

Figure 3: Header for BIGP

As shown in the above example we have two fields in the header and they are CBI (CARE BIT FOR IGP) and CBB (CARE BIT FOR BGP). When a packet enters a network where we have large number of routers, the routing information with each router is more than a limit that an IGP protocol can handle. Then the CBB bit is set to one to turn functionality of BGP "ON" as large amount of data can now be handled by this routing protocol. And in case when a packet enters a network where each router has amount of data that can be handled by an IGP protocol, then the CBI bit is set to one to turn it ON and CBB is put to zero. In this way when CBI is one it will indicate that a packet has entered an intra-autonomous communication system and thus BIGP will work as a normal IGP. And on the other hand when CBB is one it will indicate that a packet has entered an inter-autonomous kind of communication.

This header can be put on packets by a normal or any core router at the time of encapsulation. And during this time router will also ON the required care bit according to information given by its own as well its neighbour routers table (as explained in following topic).

## III. CONVERGENCE OF BIGP

We know that convergence is a very important issue when we talk of BGPs i.e. convergence rate of BGPs is slow. But all the IGPs eg. OSPF have a faster convergence. If we make convergence faster for inter-autonomous system this will make our system very unstable. Each update if flooded the way it is done in IGPs, this will crash the main backbone of our present day internet [5, 13]. So in BIGP we have a solution to this problem. We have two sets of algorithms, Algorithm1 and Alogithm2. Algorithms basically deal with setting up path of data flow and transfer of packets hence they play a very important role in handling convergence of a network under consideration. Algorithm1 is used for intra-autonomous transfer of data and basically it is similar to Dijkstra shortest path first (SPF). This algorithm calculate the now-best routes and add those to the routing TABLE A (explained in following topic) and will provide all functionalities of an algorithm used in various IGPs for path detection and transfer of data. Algorithm2 is used for inter-autonomous transfer of data and it is similar to Best path algorithm. It assigns the first valid path as the current best path and then compares the best path with the next path in the list, until BIGP reaches the end of the list of valid paths. All this data is stored in routing TABLE B. The rules that are used to determine the best path are same as Best path algorithm used in normal BGP router.

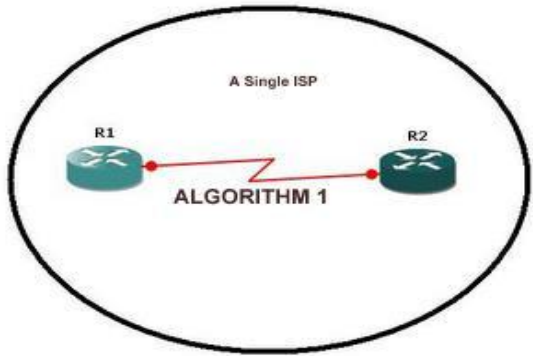

Figure 4: Intra-autonomous interaction using BIGP

When a router receives a packet having a BIGP header it checks its care bits. If CBI is ON then as shown in figure 4 router will enable its ALGORITHM1 mode so that the convergence becomes faster. And all the information will be stored in routing TABLE A of the router (explained later). The network will converge faster [20]. Timely updates after every 30-60 sec will be sent to the neighbours. Router when in this mode need not waste its CPU efficiency for taking care of routing TABLE B (explained later).

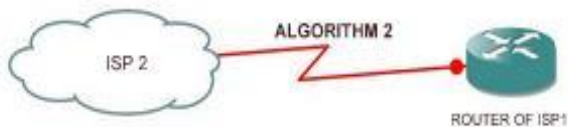

Figure 5: Inter-autonomous interaction using BIGP

When CBB bit is ON, then router switches to ALGORITHM2 mode. This makes convergence slower. The network now does not send updates at every regular interval. And routing TABLE B is filled (explained later). In this also router need not worry about routing TABLE A's entries.

## IV. SCALABILITY OF BIGP

One of the main problems with IGPs is that it is difficult to scale them whenever there is an interaction with a larger network which means large number of routers in a network. But this problem in BGP is corrected by the two ways, confederations and route reflectors [4].
But we follow a different technique in BIGP. Instead of having autonomous numbers for each autonomous system, different domain numbers in case EIGRP and OSPF, we have a single number system known as ASN for dividing inter and intra both areas of autonomous systems [7,9].

There would be a specific number range based on complexity of the network. While configuring a router we must know that if our network wants faster convergence those networks should be given a lower range of ASN and if our requirement is of lower convergence then a larger range of ASN will be given. When we receive a packet in a particular network we will first check its ASN number present in its BIGP header. If ASN number is large then there is a need of making BIGP more scalable [10].
But we can still use the concepts of confederations and route reflectors along with the use ASNs. This will make it easy to use a single protocol which can be used universally in the internet system.

## V. ROUTING TABLES OF BIGP

As discussed above the router will make entry to routing TABLE A or routing TABLE B depending on whether CBB or CBI is in ON state. But what exactly these routing tables contain is discussed in this section [12].

ROUTING TABLE A
Whenever a router receives the packet having CBI bit ON, router shifts all its CPU processing for filling this table. The basic aim is to fill entries in such a way that it makes the convergence of the network faster, neighbour relationships are made at faster rate and timely hello packets are exchanged between the routers which have enabled ALGORITHM1 mode of BIGP. So basically in this we will have the complete information of all the neighbouring routers which are in neighbour relationship with the current router. It will also have the information of the DR (designated router) and BDR (backup designated router) if elections take place in case of non-broadcasting medium. It also consists of data of entire network that is in its reach. In case our network is attached to a network using some other protocol except BIGP then a boundary router needs to be made as an ASBR, this information is also present in this table. If we have configured a smaller network as a stub or a total stub then this information should also be stored in this table. So basically all information about networks (their packets consist of ASN number in the headers) having smaller ASN values are stored in this table. This table is similar to routing table of an IGP routing protocol [17].

ROUTING TABLE B
A router commands its processor to fill this table whenever it receives a packet having CBB bit ON. All routers who enable ALGORITHM2 mode of BIGP fill this table. This happens in networks where we have smaller convergence requirement. As discussed in previous section that if in a network we have smaller convergence requirements we have larger ASNs. So basically here we store ASN (generally ASNs here have large values) values of all neighbouring routers. All the paths to reach a particular router in such case are calculated and stored [18, 19]. Here the intermediate neighbour cannot exchange the information of its other two connected neighbours in between them. For this case the other two have to be made the neighbours. So all the information related to this has to be stored in this table. In short this table is similar to routing table of a BGP routing protocol.

## VI. SIMPLE EXAMPLE OF HOW THIS NEW PROTOCOL WILL WORK

In figure 6 we have shown a simple network where we have both inter (between R3 and R4) and intra (say between R2 and R3). Suppose we need to transfer data between R3 and R4. Router R3 will send its hello packet to all its neighbours and from this it will come to know about the type of network. Since here inter-autonomous transfer of data is required R3 will set an ASN number in the BIGP header that will be selected from a larger range of ASN numbers. This ASN number will let the routers receiving the packet from R3 know that the network in which data is transferred is large and hence larger scalability is required. This will prepare other routers in ISP 1 and ISP 2 for inter-autonomous interactions. Also it will send data packets to R4 in which BIGP header will have CBB field set to one, so that BGP type of data flow can take place.

Now suppose we want to send data between R2 and R3. R2 will send hello packets to all its neighbours and from this it will come to know about type of network. Since here intra-autonomous transfer of data is required R2 will set an ASN number that is selected from smaller range of ASN numbers. This ASN number will let the routers receiving packets from R2 know that the network in which data is transferred is small. This will prepare other routers within ISP 2 for intra-autonomous interactions. Also it will send packets to R3 via R1 having BIGP header in which CBI field is set to one, so that IGP type of data flow can take place.

## VII. CONCLUSION and FUTURE WORK

In our present day internet across the globe we use different routing protocols for configuring different networks; our IBGP protocol can solve this problem. Using it we need to configure just a single routing protocol. It makes a bridge between IGP and BGP routing protocols. This can save our cost of setup and can increase the efficiency [11, 14]. Time is also saved as we need not re-distribute the packets for making data flow possible between two different routing protocols. Now we don't have to separately configure IGP and BGP. Both can be configured by configuring a single BIGP protocol. Care of scalability and convergence will be automatically done. So we can say BIGP will lead to faster network setup.

There are some limitations involved with using this kind of a protocol. It will take lots of time and effort to completely remove the concept of using different routing protocols for different types of network that are in use these days and introducing a new concept of using a single routing protocol all over the internet. Also for new protocol new specifications will need to be formulated regarding routers, wires etc. New IOS supporting this protocol need to be made and uploaded on routers so that BIGP can work on various networks.

As we have seen that we have maintained two different tables TABLE A and TABLE B. Both these tables maintain data for IGP and BGP interactions in the network respectively. If we are able to make the data of both these tables interact with each other then this will increase efficiency of BIGP hundred times. So in future we can try to make interaction between TABLE A and TABLE B possible. Also there is a need to find more ways on how to make BIGP more scalable for actual implementation in our present day internet. Also our protocol should support both IPV4 and IPV6.

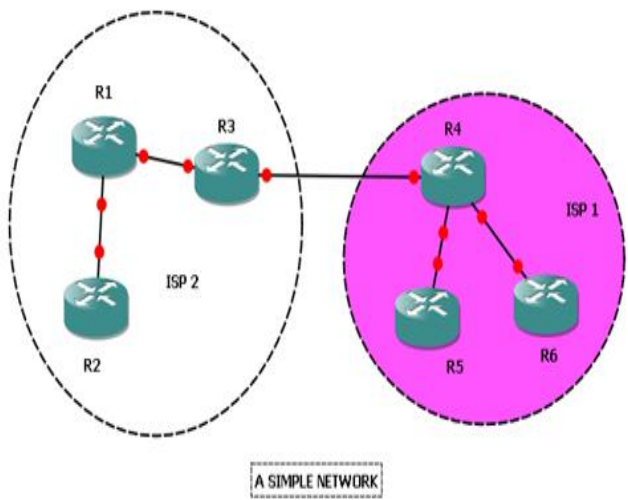

Figure 6: A Simple Network